\documentclass[%
 reprint,
 amsmath,amssymb,
 aps,
]{revtex4-2}

\usepackage{soul}% strikethrough text
\usepackage{graphicx}% Include figure files
\usepackage{float}
\usepackage{xcolor}% colored text
\usepackage{dcolumn}% Align table columns on decimal point
\usepackage{bm}% bold math
\begin{document}

\title{Directional emission in an on-chip acoustic waveguide}

\author{T.M.F. Hirsch}
\author{N.P. Mauranyapin}
\author{E. Romero}
\author{X. Jin}
\author{G. Harris}
\author{C.G. Baker}
\email{Corresponding author: c.baker3@uq.edu.au}
\author{W.P. Bowen}
\affiliation{
 ARC Centre of Excellence for Engineered Quantum Systems, School of Mathematics and Physics, University of Queensland,
St Lucia, QLD 4072, Australia}

\date{\today}% It is always \today, today,

\begin{abstract}
Integrated acoustic circuits leverage guided acoustic waves for applications ranging from radio-frequency filters to quantum state transfer, biochemical sensing and nanomechanical computing. In many applications it is desirable to have a method for unidirectional acoustic wave emission. In this work we demonstrate directional emission in an integrated single-mode, on-chip membrane waveguide, demonstrating over $99.9\%$ directional suppression and reconfigurable directionality. This avoids both loss and unwanted crosstalk, allowing the creation of more complex and compact phononic circuits.
\end{abstract}

\maketitle

% We are not SAWs. %Surface acoustic waves (SAWs) are elastic mechanical waves on the surface of a solid object~\cite{rayleighWavesPropagatedPlane1885}\textcolor{red}{Maybe we are not Rayleigh waves}. Their low loss, high spatial localisation and microwave frequency range have made them ubiquitous in electronic frequency filters~\cite{haysSurfaceacousticwaveDevicesCommunications1976}, actuators~\cite{destgeerRecentAdvancesMicrofluidic2015,dingOnchipManipulationSingle2012} and sensors~\cite{liuSurfaceAcousticWave2016}.

An acoustic integrated circuit is the mechanical-wave equivalent to an optical  or electrical circuit---a compact chip-based device that uses acoustic waves to perform a useful function~\cite{merkleinChipintegratedCoherentPhotonicphononic2017,mayorGigahertzPhononicIntegrated2021}. Phonons, the mechanical counterpart to photons and electrons, have distinct and useful properties: a high degree of spatial confinement~\cite{hatanakaPhononWaveguidesElectromechanical2014}, ultra low losses~\cite{tsaturyanUltracoherentNanomechanicalResonators2017, ghadimiRadiationInternalLoss2017,sementilliNanomechanicalDissipationStrain2022}, and robustness to electromagnetic radiation~\cite{caffeyEffectsIonizingRadiation2004,leeElectromechanicalComputing5002010}. This makes integrated acoustic circuits a subject of interest for applications such as radiofrequency filters~\cite{ruppelAcousticWaveFilter2017}, quantum-classical interfaces~\cite{bienfaitPhononmediatedQuantumState2019,goryachevObservationRayleighPhonon2013}, and biochemical sensing~\cite{arlettComparativeAdvantagesMechanical2011}.

To realise the full potential of integrated acoustic circuits for tasks such as information processing, a number of technical milestones must be achieved~\cite{fuPhononicIntegratedCircuitry2019}, including efficient acoustic driving~\cite{ekstromSurfaceAcousticWave2017a}, switching~\cite{hatanakaPhononTransistorElectromechanical2013,guerraElectrostaticallyActuatedSiliconbased2008}, and the control of acoustic interference~\cite{tadokoroHighlySensitiveImplementation2021}. A key capability to enable these milestones to be met is the ability to reconfigurably and directionally emit acoustic waves.

Directional emission is commonly achieved by emitting waves from multiple sources with different phase offsets between the sources; this has been successfully used to direct radio~\cite{yagiProjectorSharpestBeam1926}  and optical waves~\cite{hsuReviewPerspectiveOptical2021}. It has also been used to direct surface acoustic waves~\cite{collinsUnidirectionalSurfaceWave1969b}, but not in an acoustic waveguide compatible with integrated acoustic circuits. Single-mode acoustic waveguides have recently been developed~\cite{patelSingleModePhononicWire2018,romeroPropagationImagingMechanical2019}. Here we demonstrate phased emission  directly into one of these. 

Using a single mode, tightly confining waveguide obviates the need to match wave modeshapes, control the phase offset between different acoustic modes, or compensate for wave diffraction and dispersion. This reduces the need for wave emitters to have complex geometries~\cite{hodeSPUDTbasedFiltersDesign1995}, which allows fabrication of the emitters directly onto the waveguide. This in turn allows us to reduce the device footprint and avoid transduction efficiency issues related to focussing acoustic energy generated by interdigitated transducers into much smaller phononic elements~\cite{balramCoherentCouplingRadiofrequency2016,pitantiHighFrequencyMechanicalExcitation2020a,mayorGigahertzPhononicIntegrated2021,fuPhononicIntegratedCircuitry2019}. While the device we demonstrate here is a simple two-emitter phased array, the fabrication method can scale to more complex devices, such as phased arrays with more emitters or multimode waveguides capable of two-dimensional directionality. Our work paves the way for more powerful, more efficient and multipurpose integrated acoustic circuits.

\section{Method}

\begin{figure}
    \centering
    \includegraphics[width=\columnwidth]{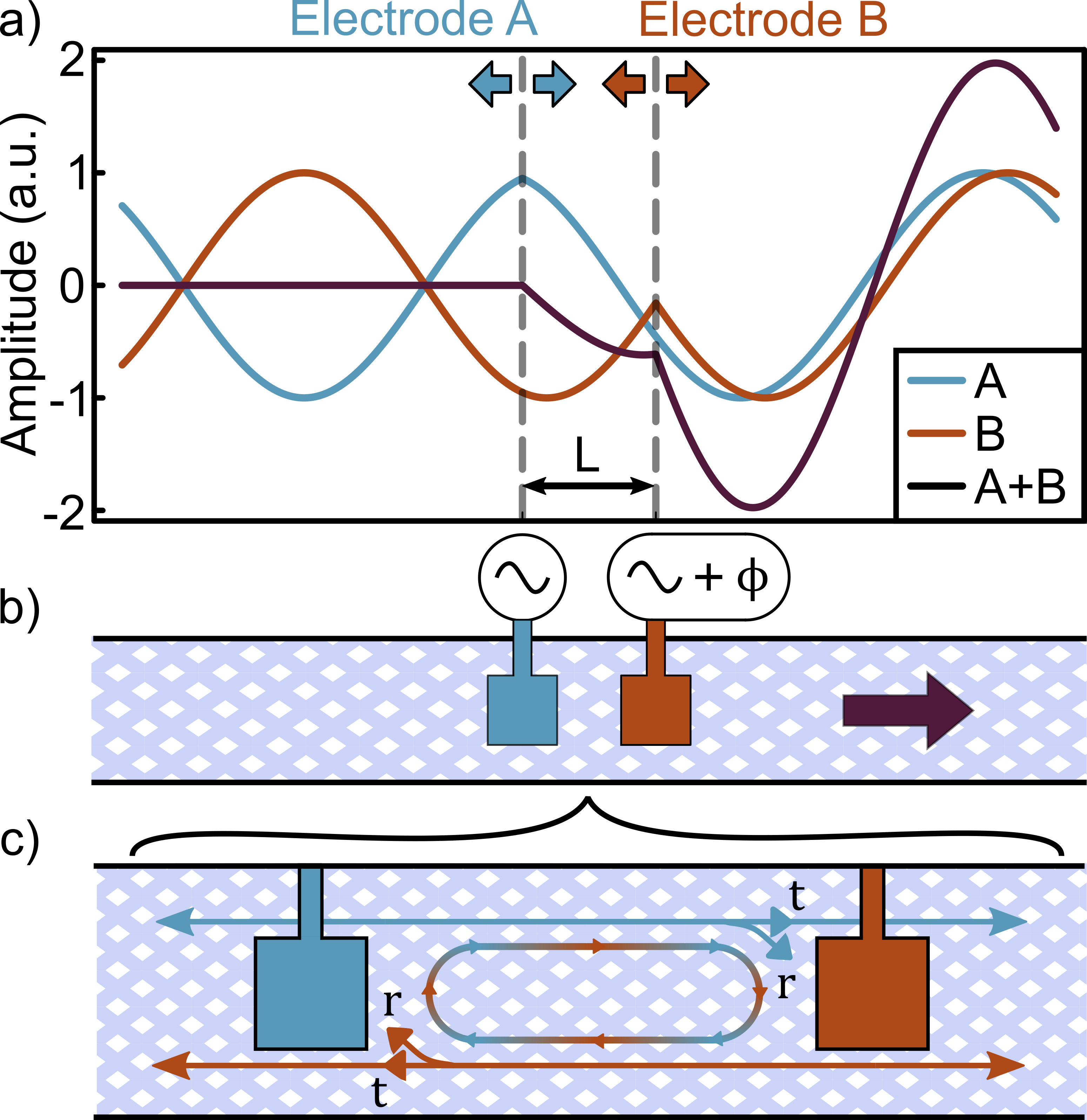}
    \caption{(a) Principle of operation for the waveguide-coupled acoustic directional emitter. Two electrodes separated by a distance $L$ act as discrete wave sources. A controllable phase difference $\phi$ between the two electrode drives is chosen such that the waves they produce constructively interfere on the right hand side and destructively interfere on the left hand side, producing unidirectional rightward power emission. Another phase difference (not illustrated) would produce constructive interference on the left hand side and destructive interference on the right hand side. (b) Schematic illustration of the physical implementation in an on-chip acoustic waveguide~\cite{mauranyapinTunnelingTransverseAcoustic2021}. (c) The electrodes introduce reflections, which we model through an amplitude reflection coefficient $r$ and transmission coefficient $t$.}
    \label{fig:principle+model}
\end{figure}

\begin{figure*}
    \centering
    \includegraphics[width=\textwidth]{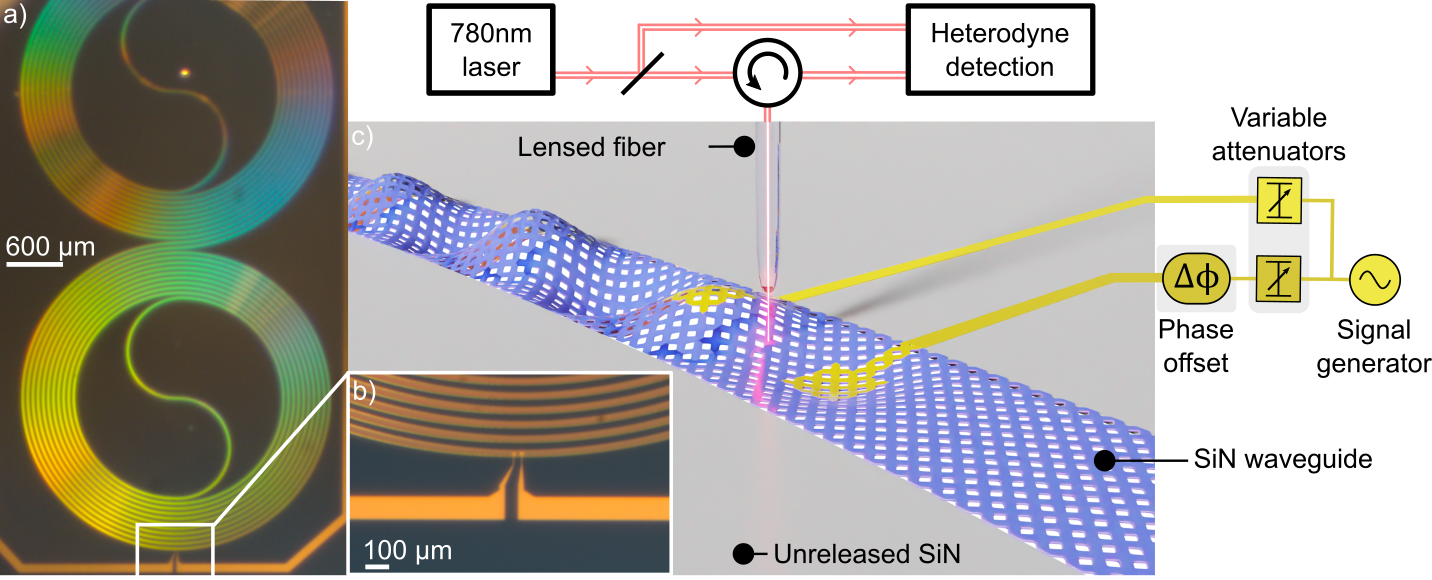}
    \caption{(a) Optical microscope image of the acoustic spiral waveguide with length 0.94 m. Iridescence arises due to the sub-micron hole pattern (not visible at this scale) used to release the membranes~\cite{mauranyapinTunnelingTransverseAcoustic2021}. (b) Close-up of the actuating electrodes. (c) 3D illustration of the experimental setup. The waveguide is a released membrane of SiN (blue). Two gold electrodes on the waveguide drive acoustic waves (not to scale). The electrodes are powered by alternating currents provided by the signal generator and tuned by variable attenuators and phase shifters. The probe laser is delivered through the lensed fiber above the waveguide.}
    \label{fig:blender+microscope}
\end{figure*}

The general principle of operation of the waveguide-coupled acoustic directional emitter is outlined in Fig.~\ref{fig:principle+model}. The waveguide is a suspended membrane of highly stressed silicon nitride that supports travelling acoustic waves~\cite{romeroPropagationImagingMechanical2019}. We use gold electrodes to capacitively actuate the waveguide~\cite{bekkerInjectionLockingElectrooptomechanical2017}. 
A natural solution for unidirectional emission might be to place a single electrode next to a waveguide terminus, but this approach is not reconfigurable and introduces resonances~\cite{hatanakaPhononWaveguidesElectromechanical2014, romeroPropagationImagingMechanical2019}. To create unidirectional emission we instead use two electrodes, A and B, located a distance $L$ apart. The total mechanical motion is given by the superposition of the wave trains produced by each electrode, as shown by the purple line in Fig.~\ref{fig:principle+model}(a).

Assuming that the electrodes do not locally affect the wave speed, which would introduce reflections, then the wave trains will constructively interfere on the left (right) side of the electrodes and destructively interfere on the right (left) side if the electrodes are driven at the same frequency with an appropriate phase difference. That difference is $\Delta\phi:=\phi_A-\phi_B=(2n-1)\pi\pm kL$, where we define the $+$ ($-$) sign to correspond to propagation in the left (right) direction, $n\in\mathbb{Z}$, and $k$ is the wavenumber. Within the single mode regime of the waveguide~\cite{romeroPropagationImagingMechanical2019}, the spatial modes of the wave trains are identical, leading to maximal interference. Choosing the correct phase difference and the electrode drive strengths to be equal results in perfect destructive interference on one side, so that the combined output will correspond to a unidirectionally propagating acoustic wave, as shown in Fig.~1(b).

In reality the electrodes introduce some level of reflection. This is caused by the acoustic impedance mismatch between the gold-coated and uncoated silicon nitride regions of the waveguide. As Fig.~\ref{fig:principle+model}(c) illustrates, the superposition is then no longer that of two wave trains, but of those two and also infinitely many reflected waves---an acoustic equivalent to the well-known Fabry-Pérot optical cavity ~\cite{hechtOpticsEugeneHecht2017,romeroScalableNanomechanicalLogic2022a}. Defining an amplitude reflection coefficient $r$ and transmission coefficient $t=\sqrt{1-r^2}$, the amplitude of waves travelling to the left direction (for example) is then:
\begin{align}\label{eqn:model}   
&u_{\leftarrow}(x,t) = \alpha e^{i(\Omega t+kx)}\biggl(V_A\nonumber\\
&+t\,\sum_{n=0}^\infty V_Be^{i(\Delta\phi)}r^{2n}e^{i(2n+1)Lk}+V_Ar^{2n+1}e^{2i(n+1)Lk}\biggr)\nonumber\\
&= \alpha e^{i(\Omega t+kx)}\left(V_L+\frac{e^{iLk}t(e^{iLk}rV_L+e^{i\Delta\phi}V_R)}{1-r^2 e^{2iLk}}\right), 
\end{align}
with the rightward travelling waves following a similar equation.
$\alpha$ is a constant of proportionality, $\Omega$ the angular frequency of the drive, $k$ the  wavenumber at frequency $\Omega$ in the waveguide, and $V_L$ ($V_R$) the voltage of the AC signal supplied to the left (right) electrode.

\begin{figure}[!ht]
    \centering
    \includegraphics[width=\columnwidth]{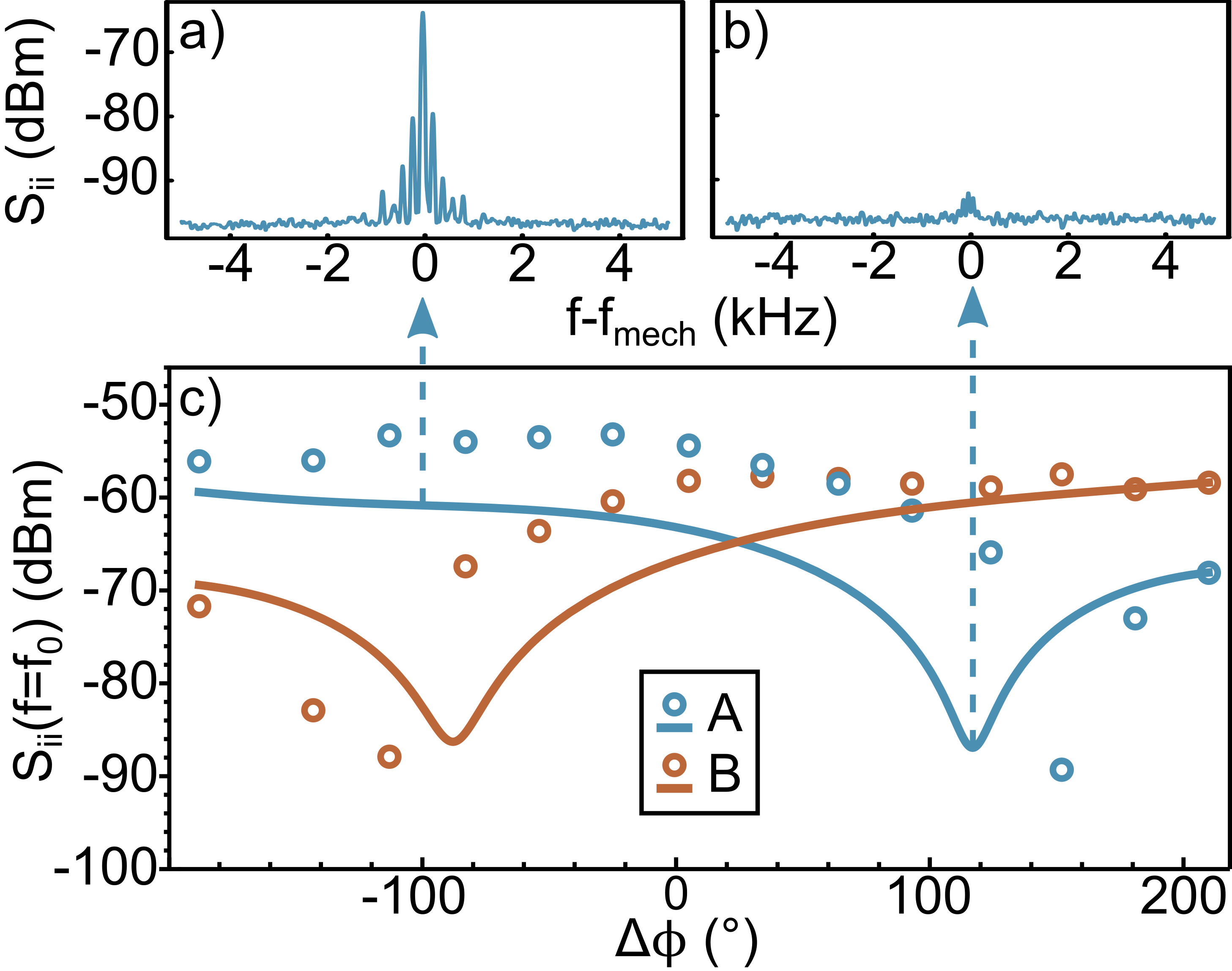}
    \caption{(a) Typical photocurrent power spectral distribution observed on a side of the electrodes when the two wave trains are  constructively interfering. $f_\mathrm{mech}=8.69\mathrm{MHz}$ is the mechanical drive frequency. (b) Typical distribution when there is destructive interference. (c) Dots: value of the photocurrent power spectral distribution at the $f=f_\mathrm{mech}-f_{LO}$, taken at fixed locations on either side of the electrodes. Solid lines: fit using Eqn.~(\ref{eqn:model}) with $r=0.9$. Because of the acoustic reflectivity discussed earlier, as the phase difference was incremented from $-188^\circ$ to $210^\circ$, the voltages to each electrode were also incremented between $200\,\mathrm{mV}$ and $1\,\mathrm{V}$.}
    \label{fig:experimental-results}
\end{figure}

We see from Eq.~\eqref{eqn:model} that reflections cause perfect destructive interference to now occur when the drive voltages are unequal---the phase difference and voltages must be chosen to zero the term in the parentheses. This can be understood from Fig.~\ref{fig:experimental-results}(b-c). The acoustic waves travelling left from electrode $B$ will partially reflect from electrode $A$ before reaching the left side, whereas leftward-travelling waves from electrode $A$ will not. As such, to achieve perfect cancellation on the left side of the waveguide, the acoustic waves coming from electrode $B$ must have higher amplitude than those coming from electrode $A$.

Figure~\ref{fig:blender+microscope} outlines the experimental setup used to demonstrate unidirectional propagation. The $30\,\mathrm{\mu m}$ wide, high tensile stress silicon nitride waveguide is released from a silicon substrate by etching through $1\,\mathrm{\mu m}$ air holes as described in~\cite{mauranyapinTunnelingTransverseAcoustic2021}. We choose to use a waveguide made out of high tensile stress silicon nitride to achieve low propagation losses for the travelling waves~\cite{tsaturyanUltracoherentNanomechanicalResonators2017}, with losses as low as $0.4\,\mathrm{dB/cm}$ as reported in~\cite{mauranyapinTunnelingTransverseAcoustic2021}. 

Generally, in acoustic waveguides, a large acoustic impedance mismatch between the membrane and the substrate is used to achieve low loss, but also causes high reflection of acoustic energy at the  extremities of the waveguide~\cite{hatanakaPhononWaveguidesElectromechanical2014, chaElectricalTuningElastic2018, romeroPropagationImagingMechanical2019}. This reflectivity means shorter acoustic waveguides in reality form high-finesse acoustic cavities, leading to the generation of standing acoustic waves which would obscure the clear demonstration of directional acoustic emission. 
%Chris had a line above: %, the magnitude of which depends on the acoustic resonance spectrum on either side of the drive electrodes, precluding
%these waveguides form acoustic cavities. In continuous operation, and demonstrate the emission of travelling acoustic waves, we We further This experiment was conducted at atmospheric pressure to increase the loss rate and reduce the wave propagation distance to further prevent resonance effects.the added contribution of air damping, This combined with the long length of the spiral waveguide,  This eliminates unwanted acoustic reflections off the ends of the waveguide, ensuring we are demonstrating purely .
To overcome this we take advantange of the high reliability of our fabrication process to fabricate a nearly meter long (94 cm) spiral delay line (Fig.~\ref{fig:blender+microscope}(a)), designed to increase the round trip time of reflected waves. On-chip delay lines like this have been fabricated for photonics~\cite{leeUltralowlossOpticalDelay2012}, however here the speed of sound is $\sim10^6$ slower than the speed of light---a useful feature for information storage and retrieval. Using the spiral delay line, combined 
with operation at atmospheric pressure, ensures that outgoing acoustic waves are fully attenuated before returning to the electrodes.

We measure mechanical motion using a heterodyne laser interferometer. A $780\,\mathrm{nm}$ probe beam generated by a Ti:Sapphire laser is focussed through a lensed fiber and reflected off the moving membrane (Fig.~\ref{fig:blender+microscope}(c)) before travelling back through the fiber as described in reference~\cite{mauranyapinTunnelingTransverseAcoustic2021}.
% CAN PUT IN TIM'S THESIS
% To see how this records mechanical motion, let the laser frequency be $\omega$, field amplitude $E_0$, light wavenumber $k$, mechanical amplitude $U_0$ and mechanical frequency $\Omega$. The reflected light was phase modulated by the motion of the membrane:
% \begin{align*}
%     E_\mathrm{refl}&=E_0 \exp(i(\omega t-2kU_0\cos\Omega t))\\
%     &=E_0\exp(i\omega t)\sum_{l=-\infty}^\infty i^lJ_l(-2kU_0)\exp(il\Omega t)\\
%     &\approx E_0\exp(i\omega t)(1+2ikU_0\sin(\Omega t)).
% \end{align*}
% The approximation uses the fact that $kU_0\ll1$, so only the first sidebands are significant. We can see that the rebounding probe light carries sidebands at $\pm\Omega$ with field amplitude proportional to the amplitude of mechanical motion.
The probe light is interfered with a local oscillator that is frequency shifted by $f_\mathrm{LO}=+80$ MHz using a fiber acousto-optic modulator in a heterodyne detection scheme. The balanced detector photo-current at $f=f_\mathrm{LO}+f_\mathrm{mech}$ is read out using a spectrum analyser.
%The photocurrent intensity is proportional to the amplitude of mechanical motion \textcolor{red}{Is this true? Not squared?}
Three stacked Smaract SLC-1720 linear piezoelectric nanopositioners allow three-dimensional positioning of the lensed fiber, and the generation of surface maps of the acoustic energy distribution, as discussed later.

The AC drive for the electrodes ($V_L$, $V_R$ in Eq. ~\eqref{eqn:model}) is provided by a signal generator driving at $f_\mathrm{mech}=8.69\,\mathrm{MHz}$. This frequency is chosen to lie above the waveguide cutoff frequency $\Omega_c=\sqrt{\sigma/\rho}(\pi/L_{wg})\sim8\,\mathrm{MHz}$, and below the frequency $\Omega_2=2\Omega_c$ at which a second transverse mode can be sustained. This allows single mode operation~\cite{romeroPropagationImagingMechanical2019}. Here $\sigma\sim1\,\mathrm{GPa}$ is the tensile stress of the silicon nitride film, $\rho\sim3200\,\mathrm{kg\cdot m}^{-3}$ the silicon nitride density, and $L_{wg}=30\,\mathrm{\mu m}$ the width of the waveguide along the transverse $y$ direction. As shown in Fig.~\ref{fig:blender+microscope}(c), the AC drive is split into two, with each branch  passing through a variable attenuator and a variable phase shifter, enabling a controllable phase offset.
%Here we measure in real time the peak-to-peak voltages and phase difference between the drives. 
Before going to the on-chip electrodes the AC drives are finally amplified and combined with a $+30\mathrm{V}_\mathrm{DC}$ bias through a bias tee to increase the capacitive drive force on the membrane~\cite{bekkerInjectionLockingElectrooptomechanical2017,schmidFundamentalsNanomechanicalResonators2016}. 

\begin{figure*}[!ht]
    \centering
    \includegraphics[width=\textwidth]{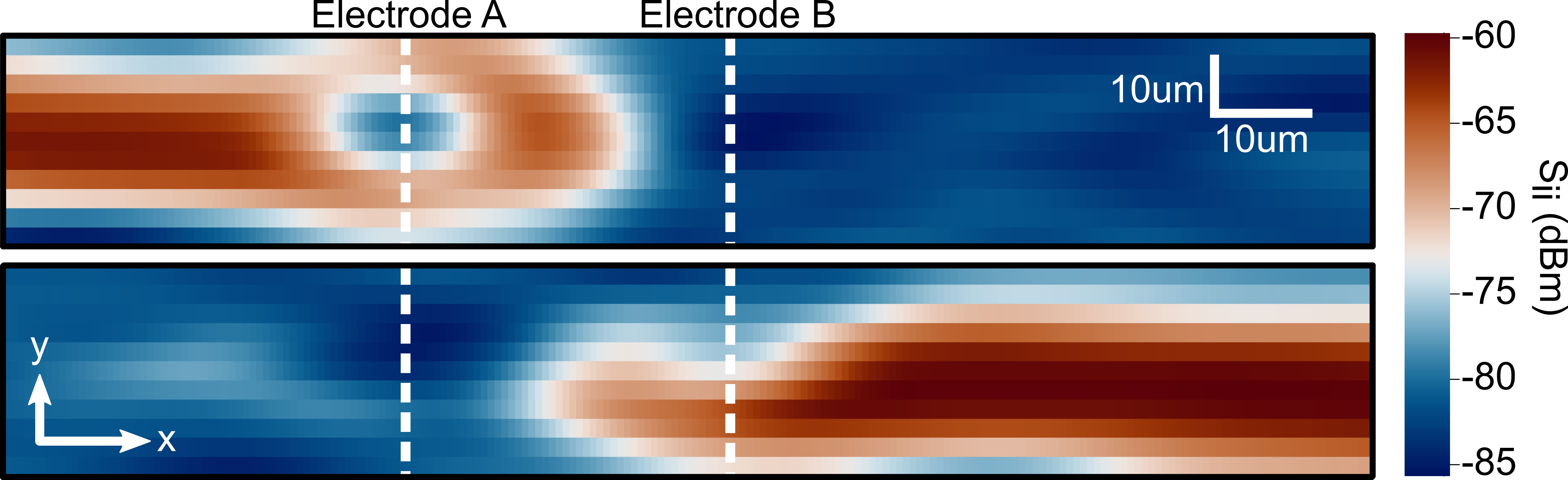}
    \caption{2D scans of the waveguide while the electrode drives were set for emission towards electrode A (top) and electrode B (bottom). Each surface plot is produced by assembling a grid of 201 by 11 measurements, covering an area of $140$ by $34$ $\mu\mathrm{m}^2$. The output is smoothed in the $x$ direction by a gaussian filter of standard deviation of 8 pixels or $\sim11\,\mu\mathrm{m}$.}
    \label{fig:2D-scans}
\end{figure*} 

To demonstrate unidirectional power emission we measure the launched acoustic power at two fixed locations, one on each side of the electrodes $\sim140\,\mathrm{\mu m}$ down the waveguide. For a given set of drive voltages and phase difference, we observe the photocurrent power spectral distribution at each location and record the photocurrent amplitude at the mechanical frequency $f=f_\mathrm{mech},$ which is to a good approximation proportional to the amplitude of mechanical motion~\cite{romeroPropagationImagingMechanical2019}. Figure~\ref{fig:experimental-results}(a) and (b) show example spectra under the conditions of constructive (a) and destructive (b) interference. To verify that the acoustic energy is indeed being controllably routed to either side of the electrodes, we perform a systematic sweep of the phase difference $\Delta \phi$, at each setting recording the photocurrent amplitude at both locations.
The result of this measurement is shown in Fig.~\ref{fig:experimental-results}(c), where the photocurrent amplitude on the left (right) of the electrodes is plotted by the blue (red) open circles, respectively. This confirms that an emission maximum on one side is accompanied by a minimum on the other, as expected. 
%We performed two experiments, the results of which are presented in Figures \ref{fig:experimental-results} and \ref{fig:2D-scans}.
We overlay on top of this a fit based on the model in Eq.~(\ref{eqn:model}) (solid lines). This provides qualitative agreement with the experimental data, capturing the main features of the process. Some discrepancy is present, notably in terms of the phase offset $\Delta \phi$ of the minima, which are $20\%$ greater experimentally than predicted. We attribute this to several possible mechanisms. First, the model neglects frequency drift, the nonzero physical size of the electrodes, and the excitation of higher order evanescent modes in the immediate vicinity of the electrodes~\cite{romeroPropagationImagingMechanical2019}. Second, there is also the presence of Duffing nonlinearity~\cite{romeroScalableNanomechanicalLogic2022a} in the waveguide region between the electrodes where the wave amplitude is enhanced. Duffing nonlinearity causes an amplitude-dependent speed of sound, leading to an additional phase shift which isn't taken into account in Eq.~(\ref{eqn:model}). We verify the presence of Duffing nonlinearity by configuring the electrodes for directional suppression, then varying $V_1$ and $V_2$ while keeping their ratio $V_1/V_2$ constant, and observing that the level of directional suppression is reduced.

Our measurements demonstrate highly efficient waveguide-coupled directional power emission.  The configuration at $\Delta\phi=152^\circ$ corresponds to a 31.8~dB power difference between the left and right sides, that is $99.93\%$ of the acoustic power is emitted to the left side of the electrodes. Similarly, the configuration at $\Delta\phi=113^\circ$ corresponds to acoustic power steered to the right side with even higher 34.6~ dB = $99.96\%$ selectivity.

%The fit of the model, provided by Eq \eqref{eqn:model}, shows we have a good understanding. However, it does predict slightly closer together minima. We attribute this to the presence of Duffing nonlinearity~\cite{romeroScalableNanomechanicalLogic2022a} in the waveguide in the region between the electrodes where the wave amplitude is enhanced. Duffing nonlinearity causes an amplitude dependent speed of sound, leading to an additional phase shift which would interfere with the superposition.

To further confirm this ability to controllably steer the acoustic energy in either direction of the waveguide, we produce 2D surface plots of the acoustic wave emission. We achieve this by rastering the lensed fiber in both $x$ and $y$ directions~\cite{mauranyapinTunnelingTransverseAcoustic2021}, while measuring the acoustic power at the drive frequency. These measurements are shown in Fig.~\ref{fig:2D-scans}. We first scan with the electrode drives tuned to produce destructive interference on the right side and acoustic emission toward the left (top panel of Fig.~\ref{fig:2D-scans}). Next, the drives are tuned for acoustic emission in the other direction (bottom panel of Fig.~\ref{fig:2D-scans}).

Figure~\ref{fig:2D-scans} shows clear evidence that the directional emitter works, with  $>99$\% directional suppression. Out of view of the scan the waves exponentially decay because of the significant squeeze-film air damping that occurs at room pressure~\cite{parrainDampingOptomechanicalDisks2012}.
The nearly uniform power distribution at this scale along the length $x$ of the waveguide---with an absence of nodes and antinodes---is a confirmation of the absence of acoustic interference, and demonstrates true directional emission of travelling acoustic waves in an integrated waveguide.

We have demonstrated directional emission, achieving 30~dB of suppression and easy reconfigurability. The device demonstrated is a simple concept using just two electrodes.
%Future improvements: reduce reflection by redesigning the gold electrodes to reduce the sharpness of the impedance gradient. Change spacing to nearer to $\lambda/4$ of the mechanical wavelength, which is the theoretically optimal distance where perfect constructive interference coincides at the same phase difference as perfect destructive interference.
This could be straightforwardly expanded to achieve 2D directional control by using a multimode waveguide and adding more electrodes in an acoustic phased array. The ability to generate directional emission in on-chip acoustic waveguides, and to control this dynamically, could be used in novel phononic circuits, for example to selectively probe multiple devices distributed over a macroscopic membrane resonator~\cite{chienSinglemoleculeOpticalAbsorption2018}, deliver pulses of acoustic energy for highly confined nonlinear behaviour~\cite{kurosuOnchipTemporalFocusing2018b}, or---combined with the accessible nonlinearity and tunable coupling of the membrane platform~\cite{mauranyapinTunnelingTransverseAcoustic2021}---actuate the nodes of a mechanical computer~\cite{romeroScalableNanomechanicalLogic2022a}.

% Chris 29/9/23: another use would be sending individual pulses in different directions at different times, to combine and perform nonlinear functions e.g. 4WM or SPM at a desired location [cite the Hatanaka pulse chirping paper, 4WM paper]

\begin{acknowledgments}
This research was primarily funded by the Australian Research Council and the Lockheed Martin Corporation through the Australian Research Council Linkage Grant No. LP190101159. Support was also provided by the Australian Research Council Centre of Excellence for Engineered Quantum Systems (No. CE170100009). G.I.H. and C.G.B. acknowledge their Australian Research Council grants (No. DE210100848 and No. DE190100318).
This work was performed in part at the Queensland node of the Australian National Fabrication Facility and the Australian Microscopy \& Microanalysis Research Facility at the Centre for Microscopy and Microanalysis.
\end{acknowledgments}

\begin{center}
    \small{\textbf{DATA AVAILABILITY}}
\end{center}

The data that support the findings of this study are available from the corresponding author upon reasonable request. % APL statement: https://publishing.aip.org/resources/researchers/author-instructions/#data

\bibliography{bib}% Produces the bibliography via BibTeX.

\end{document}